\documentclass[journal]{IEEEtaes}
\usepackage{amsmath,amssymb,amsfonts}
\usepackage{graphicx}
\graphicspath{{./figs/}}
\usepackage{bm}
\usepackage[numbers,sort&compress]{natbib}
\usepackage{balance}
\usepackage{comment}
\usepackage{xcolor}
\usepackage{colortbl}
\usepackage{url}
\usepackage{mathrsfs}
\usepackage{bbm}
\usepackage{hyperref}
\usepackage[acronym]{glossaries}
\glsdisablehyper
\usepackage[T1]{fontenc}
\usepackage[utf8]{inputenc}
\usepackage{cleveref}
\labelformat{subsection}{\thesection-#1}
\usepackage{subcaption}
\usepackage{tabularx}
\usepackage{array}
\usepackage{diagbox}
\newcolumntype{C}{>{\centering\arraybackslash}X}
\newcolumntype{L}[1]{>{\raggedright\arraybackslash}p{#1}}
\usepackage{float}
\usepackage{diagbox}
\usepackage[linesnumbered,ruled,vlined]{algorithm2e}
\usepackage{adjustbox}
\usepackage{booktabs} 
\jvol{}
\jnum{}
\jmonth{}
\paper{}
\pubyear{}
\doiinfo{}
\doiinfo{}
\newacronym{cir}{CIR}{channel impulse response}
\newacronym{cfar}{CFAR}{constant-false-alarm rate}
\newacronym{trbd}{TrBD}{track-before-detect}
\newacronym{snr}{SNR}{signal-to-noise ratio}
\newacronym{sinr}{SINR}{signal-to-interference-plus-noise ratio}
\newacronym{inr}{INR}{interference-to-noise ratio}
\newacronym{ekf}{EKF}{extended Kalman filter}
\newacronym{gospa}{GOSPA}{generalized optimal subpattern assignment}

\newcommand{\qgzero}[1]{\cellcolor{black!30}#1} 
\newcommand{\qgone}[1]{\cellcolor{black!25}#1}  
\newcommand{\qgtwo}[1]{\cellcolor{black!20}#1}  
\newcommand{\qgthree}[1]{\cellcolor{black!15}#1}
\newcommand{\qgfour}[1]{\cellcolor{black!10}#1} 

\definecolor{qblue}{RGB}{207,226,243}
\definecolor{qblueviolet}{RGB}{194,211,240}
\definecolor{qviolet}{RGB}{190,203,238}
\definecolor{qpurple}{RGB}{207,187,226}
\definecolor{qpink}{RGB}{235,177,189}
\definecolor{qred}{RGB}{248,154,154}

\newcommand{\qfive}[1]{\cellcolor{qblue}#1}          
\newcommand{\qsix}[1]{\cellcolor{qblueviolet}#1}    
\newcommand{\qeight}[1]{\cellcolor{qpurple}#1}      
\newcommand{\qnine}[1]{\cellcolor{qpink}#1}         
\newcommand{\qhigh}[1]{\cellcolor{qred}\textbf{#1}} 

\begin{document}

\title{A Computationally Efficient Likelihood Approximation for Target Tracking in Time-Varying Multipath Channels}

\author{Ashwani Koul}
\member{Student Member, IEEE}
\affil{{}Linköping University, Linköping, Sweden} 

\author{Gustaf Hendeby}
\member{Senior Member, IEEE}
\affil{{}Linköping University, Linköping, Sweden}

\author{Isaac Skog}
\member{Senior Member, IEEE}
\affil{KTH, Stockholm, Sweden, and\\ 
FOI-Swedish Defence Research Agency, Kista, Sweden}


\receiveddate{
This work was partially supported by the Wallenberg AI, Autonomous Systems and Software Program (WASP), funded by the Knut and Alice Wallenberg Foundation.}

\corresp{}

\authoraddress{Ashwani Koul and Gustaf Hendeby are with Division of Automatic Control, Department of Electrical Engineering, Linköping University, Linköping, Sweden-581 83, Email: (ashwani.koul@liu.se, gustaf.hendeby@liu.se); Isaac Skog is with Department of Electrical Engineering, KTH, Sweden-100 44, and Dept. of Underwater Technology, FOI-Swedish Defence Research Agency, Kista, Sweden, Email: (skog@kth.se)}

\editor{}
\supplementary{}

\markboth{KOUL ET AL.}{Background Aware TrBD filtering}
\maketitle

\begin{abstract}
Active sonar target tracking in shallow-water environments is challenging when weak target echoes are embedded in a time-varying background containing structured multipath components. Conventional detect-before-track methods rely on thresholded detections, which may discard weak target evidence or generate false tracks from multipath-induced detections. At low signal-to-noise ratios, track-before-detect filtering can improve tracking performance by exploiting weak target information directly from raw sensor measurements, but directly accounting for the time-varying background leads to a joint target--background inference problem, which is computationally demanding. To address this, this paper develops a physics-motivated and computationally efficient approximation of the raw sensor measurement likelihood for Bernoulli track-before-detect filtering. The multipath components in the background are modeled in the raw sensor measurement domain and recursively tracked using an extended Kalman filter. By neglecting the posterior dependence between the target and background state, the predicted background statistics are used to construct approximate target-present and target-absent likelihoods for the Bernoulli filter. Evaluations using measurements generated from the statistical model and BELLHOP show improved target-confirmation and localization performance over constant-false-alarm-rate-based tracking. These results indicate that explicitly accounting for the background through a computationally tractable approximate likelihood can exploit weak target information without requiring full joint target--background inference.
\end{abstract}


\section{INTRODUCTION}
Accurate modeling of sensor measurements is fundamental to reliable detection, localization, and tracking. The performance of any tracking algorithm depends not only on the choice of filter or data-association method, but also on the statistical model used to describe the measurements~\cite{Abraham2019, BarShalom2011}. If the measurement model does not capture the dominant physical effects in the sensing environment, the resulting likelihood may be mismatched to the sensor measurements, leading to missed detections, false alarms, biased localization estimates, and unstable tracks.

\vspace{1.5pt}
This issue of likelihood mismatch is particularly important in active sonar, where the weak target echo is embedded in a background consisting of ambient noise and structured, correlated components generated by time-varying multipath propagation and reverberation. Consequently, an independent and identically distributed (i.i.d.) Gaussian-noise model is often inadequate in such scenarios~\cite{Abraham2019}. This can cause target echoes to be masked by unmodeled background components, making reliable target detection and tracking difficult.

\vspace{1.5pt}
A common approach to target detection and tracking is the classical detect-before-track framework, in which the raw sensor measurements are first transformed into representations such as delay--Doppler maps or bearing--time records. A \gls{cfar} detector is then applied to these representations to generate point detections, which are subsequently provided to a tracking algorithm~\cite{BarShalom2011, Blackman1999, Richards2022, Abraham2019}. The \gls{cfar}-based detect-before-track methods are generally most effective at moderate-to-high \gls{snr}, where the local background estimates within the CFAR detector can be used to reliably set the adaptive detection threshold. However, at low \gls{snr}, weak target returns may fall below the detection threshold, while multipath components in the background may generate false detections, resulting in  false tracks.

\vspace{1.5pt}
\Gls{trbd} methods address the problem of detecting and tracking targets in low-\gls{snr} environments~\cite{Ristic2013, Wang2019, Buzzi2008, Zhang2021}. A \gls{trbd} filter processes raw sensor measurements or the unthresholded transformed measurements and accumulates weak target evidence across multiple scans or pings using a model of the target motion. This capability comes at the cost of increased computational cost and greater sensitivity to modeling errors, both in terms of likelihood function and target motion model. Rather than making a hard detection decision at each ping, a \gls{trbd} filter uses the measurement likelihood to update both the target-existence probability and the target-state posterior distribution recursively. 

\vspace{1.5pt}
Modeling the statistical distribution of the measurements directly in the raw measurement domain provides a natural framework for target tracking in time-varying multipath channels. In this way, the model parameters remain connected to the underlying propagation physics~\cite{Yang2012, Walree2013, Huang2013a}. However, the exact Bayesian inference problem becomes computationally demanding as the target state and the background need to be estimated jointly. Many existing sonar tracking methods reduce the computational load by operating on point detections extracted from delay--Doppler maps or bearing--time records~\cite{Li2015, Li2021, Yang2024a, Kim2024}. Although the amount of data processed by the tracker is reduced, the statistical characterization of the time-varying background becomes more difficult. Consequently, multipath components in the background may generate false detections, resulting in false tracks.

\vspace{1.5pt}
Several studies have addressed background inference for target-detection applications. In~\cite{Yang2024}, block updating of a sparse and structured \gls{cir} model is employed for weak-target detection. In~\cite{Jia2021}, nonnegative matrix factorization is used to infer the reverberation in the background, whereas low-rank and sparse decomposition is employed for reverberation reduction in~\cite{Zhu2022}. These studies demonstrate the usefulness of explicitly modeling the structured background, but these methods have not been developed to construct a raw-measurement likelihood suitable for target-state inference within a \gls{trbd} filter.

\vspace{1.5pt}
To address this limitation, this paper extends the background-tracking framework proposed in~\cite{Koul2026} to active-sonar target tracking. In~\cite{Koul2026}, the time-varying multipath channel is modeled in the raw sensor measurement domain using a wideband Doppler linearization, and the background is recursively tracked with an \gls{ekf}. In the proposed method, the joint target--background inference problem is made computationally efficient by neglecting the posterior dependence between the target and background state, allowing the background state to be inferred separately. The predicted background statistics are then used to construct an approximate measurement likelihood for target inference. By combining this physics-motivated background model with direct processing of the raw sensor measurements, the proposed framework adapts the measurement likelihood to the time-varying multipath channel while preserving weak target information at low \gls{snr}.
 
 \vspace{1.5pt}
To summarize, the main contributions of this paper are as follows
\vspace{-0.55\baselineskip}
\begin{itemize}
\item a computationally efficient, physics-motivated likelihood formulation based on raw sensor measurements for \gls{trbd} filtering in active sonar.
\item a low-\gls{snr} evaluation under background-model mismatch conditions using BELLHOP-generated time-varying multipath signal propagation.
\end{itemize}
\vspace{-1\baselineskip}
\noindent \textbf{Reproducible research}: The code and data used to reproduce the results are available at:\\ 
$\texttt{\url{https://github.com/ASHKoul/Bcg_SLRT}}$

\section{PROBLEM FORMULATION}
For accurate target detection and tracking, a reliable model of the likelihood ratio is required. The likelihood ratio is defined as
\begin{equation}
    L \triangleq \frac{\ell_1}{\ell_0}
\end{equation}
where  $\ell_0$ and $\ell_1$ denote the likelihood functions under the target-absent and target-present hypotheses, respectively. Here, the target-absent and target-present hypotheses are defined as
\begin{subequations}
\begin{align}
\mathcal{H}_0 &: \text{target absent}, \\
\mathcal{H}_1 &: \text{target present}. 
\end{align}
\end{subequations}

\begin{figure}[t]
    \centering
    \includegraphics[width=0.5\textwidth, height=0.25\textwidth]{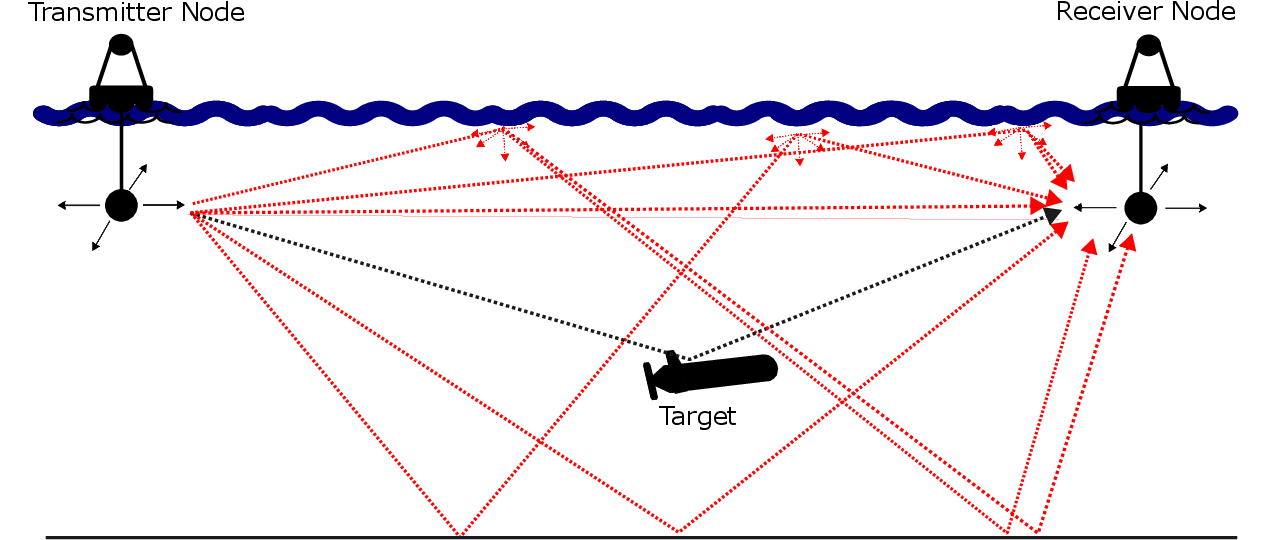}
    \caption{Illustration of the considered active sonar scenario in which the received measurements contain the target echo and the effects of a time-varying multipath channel.}
    \label{fig:channel}
\end{figure}

For the active sonar scenario shown in Fig.~\ref{fig:channel}, the received signal at receiver $j$ is modeled as
\begin{equation}\label{eq: sub_eq_measu_parts}
y_j(t)=x_j^b(t)+x_j^o(t)+e_j(t),
\end{equation}
where $x_j^b(t)$ is the time-varying multipath component of the background, $x_j^o(t)$ is the target return, and $e_j(t)$ denotes the ambient noise. Under hypothesis $\mathcal{H}_0$, $x_j^o(t)=0$, whereas under hypothesis $\mathcal{H}_1$, $x_j^o(t)\neq0$. Together with the assumed statistical model for the ambient noise, the measurement model defines the statistical distributions of the received measurements under both hypotheses, which form the basis for the likelihood functions used for target detection and tracking.

Since both $x_j^b(t)$ and $x_j^o(t)$ are generated from delayed and Doppler-scaled versions of the transmitted waveform, a weak target echo may be difficult to distinguish from the time-varying multipath components. Consequently, an inaccurate background model may result in mismatched likelihood functions and degraded detection and tracking performance. Moreover, the joint inference of the target and background state is computationally demanding. The proposed formulation addresses these challenges by introducing a set of approximations and neglecting the posterior dependence between the target and background state. This allows the background state to be estimated separately using an \gls{ekf}. The predicted background statistics are subsequently used to approximate the $\ell_0$ and $\ell_1$ likelihoods, and the likelihood ratio $L$, for \gls{trbd} filtering. 

\section{MEASUREMENT MODEL AND BACKGROUND TRACKING}\label{sec: backgrnd_Model}
This section develops a statistical model for the time-varying background and a recursive method for tracking its statistics, which are subsequently used to construct the approximate measurement likelihood. The objective is not to resolve every physical propagation path individually, but to capture the dominant delay and Doppler variations required to accurately approximate the raw sensor measurement likelihood.

\subsection{Multipath Propagation Model}
Assuming that the propagation parameters remain approximately constant for the time duration $T$ then $x^{b}_{j}(t)$ in~\eqref{eq: sub_eq_measu_parts} can be written as~\cite{Liu2012}
\begin{equation}\label{eq: sub_eq_multipath}
    x^{b}_{j}(t)=\sum_{i=1}^{N_b} a_{i,j} s(\beta_{i,j}(t-\tau_{i,j})) \qquad t\in [0,\,T],
\end{equation}
where $s(t)$ is the transmitted waveform. Further, $\tau_{i,j}$, $\beta_{i,j}$, and $a_{i,j}$ are the delay, Doppler scaling, and amplitude of the $i$:th arrival, respectively. Here, $N_b$ is the associated number of multipaths in the background. For Doppler-scale factors close to unity, i.e., $\beta\approx1$, the multipath components can be approximated using a wideband Doppler linearization (see Appendix~\ref{app:linearization}) as  
\begin{align}\label{eq: wideband approximation}
x^{b}_{j}(t)\approx\; \sum_{i=1}^{N_b} & a_{i,j}\, s(t-\tau_{i,j}) \notag\\
   + \sum_{i=1}^{N_b}& a_{i,j}\,r_{i,j}\,(t-\tau_{i,j})\dot s(t-\tau_{i,j}), \notag\\
\approx\;\sum_{i=1}^{N_b} & \Big( a_{i,j}\, s(t-\tau_{i,j}) +a_{i,j}\,r_{i,j}\,u(t-\tau_{i,j}) \Big),
\end{align}
where  $r_{i,j} \triangleq \ln \beta_{i,j}$ denotes the log-Doppler-scale parameter, and $ u(t) \triangleq t\,\dot{s}(t)$. 

\subsection{Multipath Approximation for the Target}
The received target signal $x_j^o(t)$ may also contain several propagation paths. To maintain a tractable target-state-dependent likelihood, only the dominant bistatic target-return path is retained, while weaker target-generated multipath components are neglected~\cite{Urick2013, Waite2002}. Under this approximation, the target return at receiver $j$ is modeled as
\begin{align}\label{eq: target}
    x^{o}_{j}(t) & \approx a_{o,j} s(\beta_{o,j}(t-\tau_{o,j})).
\end{align}
Here  $\tau_{o,j}$, $\beta_{o,j}$, and $a_{o,j}$ denote the delay, Doppler-scale factor, and amplitude of the dominant target return, respectively.

\subsection{Discrete-Time Measurement Model}
Substituting the background approximation in \eqref{eq: wideband approximation} and the target model in \eqref{eq: target} into~\eqref{eq: sub_eq_measu_parts} gives 
\begin{equation}\label{eq: meas_final_cont_time}
\begin{split}
    y_j(t) &  \approx \sum_{i=1}^{N_b} \Big( a_{i,j}\, s(t-\tau_{i,j})\;+\; a_{i,j}\,r_{i,j}\,u(t-\tau_{i,j}) \Big)\\
        &  \qquad + \,\, a_{o,j} s(\beta_{o,j}(t-\tau_{o,j})) +e_{j}(t).
\end{split}
\end{equation}
Next, let $T_s$ be the sampling period and define
\begin{equation}\label{eq: sampled signal}
s[n]\triangleq s(nT_s) \quad \textrm{and}\quad u[n]\triangleq nT_s\,\dot s(nT_s)
\end{equation}
with the corresponding zero-padded Toeplitz matrices as
\begin{equation}
[S]_{n,l}\triangleq s[n-l] \quad \textrm{and}\quad [U]_{n,l}\triangleq u[n-l],
\end{equation}
where $l=0,\cdots, N_l-1$ represents the sampled delay grid with $ N_l$ as the grid length, and samples outside the waveform support are set to zero. Moreover, let $y_{j,k}$ represent the discrete-time measurement vector defined over $N\triangleq \lceil T/T_s\rceil$ samples collected during $k$:th ping 
\begin{equation}
    y_{j,k}=\begin{bmatrix} y_{j,k}[0] & \ldots & y_{j,k}[N-1] \end{bmatrix}^\top.
\end{equation} 
Then, a discrete-time measurement model for $y_{j,k}$ is given by
\begin{equation}\label{eq: complete_disc_meas_equ}
    y_{j,k}= Sa_{j,k} + U \Lambda_{a_{j,k}} r_{j,k} + x^o_{j,k} + e_{j,k},
\end{equation}
where $$a_{j,k}=\begin{bmatrix} a_{j,k}[0] & \ldots & a_{j,k}[N_l-1] \end{bmatrix}^\top,$$ and $$r_{j,k}=\begin{bmatrix} r_{j,k}[0] & \ldots & r_{j,k}[N_l-1]\end{bmatrix}^\top$$ are $N_l$ element vectors collecting the multipath amplitudes and log-Doppler-scale parameters, respectively. 
Further, $\Lambda_{a_{j,k}}\triangleq\mathrm{diag}(a_{j,k})$, and $e_{j,k}$ represents the discrete-time ambient measurement noise and residual modeling errors. Further, the element $r_{j,k}[\,l\,]$ is the log-Doppler-scale parameter associated with the $l$:th delay tap index.

\subsection{Statistical Channel Model}
\begin{figure}[t]
    \centering
    \includegraphics[width=0.5\textwidth,height=0.35\textwidth]{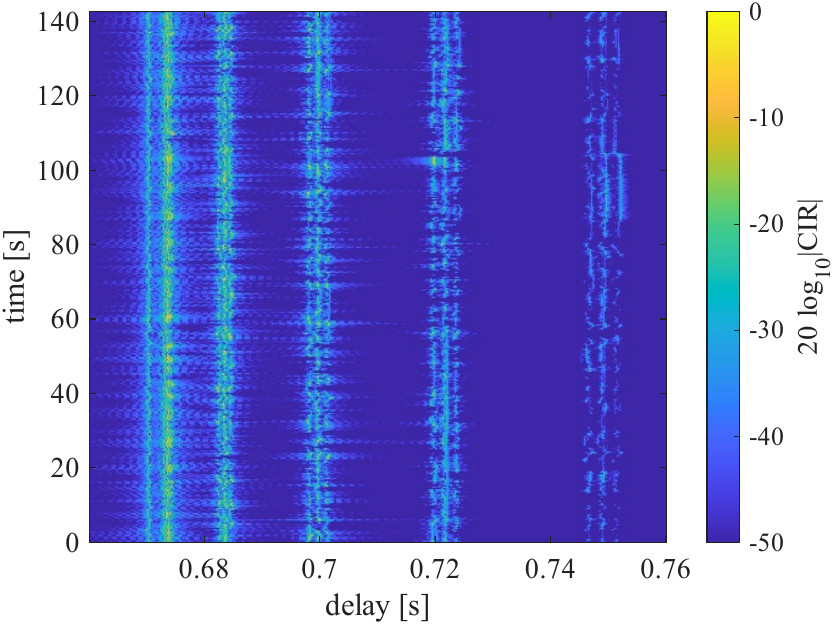}
    \caption{Simulated \gls{cir} using BELLHOP.}
    \label{fig: CIR}
\end{figure}

The moving ocean surface and small relative motion of the transmitter and receiver induce temporal variations in the background. These variations may contain both path-specific and common components. Local scattering and path-dependent boundary interactions produce path-specific perturbations, whereas platform motion and broader environmental changes can affect several propagation paths simultaneously~\cite{miliucom}. The resulting channel therefore exhibits temporal variation within individual paths together with statistical dependence across paths. The BELLHOP-simulated \gls{cir} in Fig.~\ref{fig: CIR} illustrates the persistent multipath structure and its temporal variation.

To capture these effects, the Doppler-induced random fluctuations are statistically modeled as
\begin{equation}
    r_{j,k} \triangleq c_{j,k} + d_{j,k},
\end{equation}
where $ c_{j,k}\sim \mathcal{N}(0,\sigma_{c,j}^2I)$ represents independent path-specific perturbations and $d_{j,k}\sim \mathcal{N}(0,\sigma^2_{d,j}\mathbbm{11^\top})$ represents a common perturbation shared by all delay taps. Here, $\mathbbm{1}$ is a vector with all values equal to 1. This specific decomposition captures both independent path-dependent variability and correlated common-mode perturbations. The effectiveness of such a modeling structure has been shown in~\cite{Koul2026} for background tracking and target detection. 

Next, assuming the ambient noise $e_{j,k}$ is white and normally distributed with known variance $\sigma^2_e$, the distribution of the measurement vector conditioned on the channel amplitudes and the target signal is given by
\begin{equation}\label{eq: prob_dist_measu}
    p(y_{j,k}\mid a_{j,k},x^o_{j,k})= \mathcal{N}\big(y_{j,k};Sa_{j,k} +x^{o}_{j,k},\;R(a_{j,k})\big)
\end{equation}
where $$R(a_{j,k})\triangleq \sigma_e^2 I + \Sigma_b(a_{j,k}),$$  
with
\begin{equation}\label{eq:Sigma_b_a}
\Sigma_b(a_{j,k})= \Sigma_c (a_{j,k}) +\Sigma_d(a_{j,k}).
\end{equation}
Here, 
\begin{equation}\label{eq:Sigma_c_d}
\begin{aligned}
\Sigma_c(a_{j,k}) \triangleq & \,\,\sigma_{c,j}^2 \,U \Lambda_{a_{j,k}}\Lambda_{a_{j,k}}^\top U^\top,\\
 \Sigma_d(a_{j,k}) \triangleq & \,\,\sigma_{d,j}^2 \,U\,a_{j,k}\,a_{j,k}^\top\, U^\top.
\end{aligned}
\end{equation}

\subsection{Low-Dimensional Representation}
The received signal may be dominated by diffuse reverberation for multipath arrivals with longer propagation delays. The scattering from the ocean boundaries and inhomogeneities within the water column spreads the received energy over neighboring delays and induces smoothness across the delay taps. To exploit this structure and avoid an unnecessarily high-dimensional background state vector, the delay-domain amplitude vector is represented using a fixed basis matrix, i.e.,
\begin{equation}\label{eq:a_basis_theta}
   a_{j,k} =  B \theta_{j,k},
\end{equation}
where $\theta_{j,k}$ is an $M \times 1$ vector of basis coefficients and $B \in \mathbb{R}^{N_l\times M}$ is constructed using Gaussian radial basis function,
\begin{equation}
[B]_{l,m}\triangleq \exp\!\left(-\frac{(lT_s-\mu_m)^2}{2\sigma_m^2}\right),
\end{equation}
where $\mu_m$ and $\sigma_m$ are the center and length scale of the $m$:th basis, respectively. The basis centers are uniformly spaced over the considered delay interval with spacing $\Delta\mu=1/\mathrm{BW}$, and the length scale is set to $\sigma_m=1/\mathrm{BW}$. Now, given~\eqref{eq:a_basis_theta}, the conditional distribution in \eqref{eq: prob_dist_measu} becomes
\begin{equation}\label{eq:reparam_stat_model_clean}
 p(y_{j,k}\mid \theta_{j,k},x^o_{j,k})= \mathcal{N}\!\big(y_{j,k}; H\theta_{j,k} +x^{o}_{j,k},\;R(\theta_{j,k})\big),
\end{equation}
with $H\triangleq SB$.

\subsection{Background Tracking}
Let $\theta_{j,k}$ denote the background state vector at receiver $j$ during ping $k$. The background state vector $\theta_{j,k}$ is assumed to evolve gradually between consecutive pings and is modeled as a first-order Markov process,
\begin{equation}
p(\theta_{j,k+1}\mid\theta_{j,k})=\mathcal{N}\left(\theta_{j,k+1};\theta_{j,k},Q_{\theta_j}\right),
\label{eq:back_dyn_model}
\end{equation}
where the process-noise covariance is chosen as
\begin{equation}
[Q_{\theta_j}]_{m,n}=\left(\frac{\sigma_{w,j}}{m}\right)^2\delta_{m,n}, \quad \forall \, m=1,\ldots,M
\end{equation}
and $\delta_{m,n}$ denotes the Kronecker delta. The process-noise variance is assumed to decrease with increasing basis index $m$. This is based on the assumption that later delay components, which typically undergo more surface and bottom interactions, exhibit smaller temporal variations than the earlier arrivals.

Now, to track the background, the measurement model is evaluated under the target-absent hypothesis $\mathcal{H}_0$. Setting $x^{o}_{j,k}=0$ in~\eqref{eq:reparam_stat_model_clean} gives
\begin{equation}
p(y_{j,k}\mid\theta_{j,k};\mathcal{H}_0)=\mathcal{N}\left(y_{j,k};H\theta_{j,k},R(\theta_{j,k})\right).
\label{eq:bg_meas_model}
\end{equation}
The state-transition model in~\eqref{eq:back_dyn_model} together with the measurement model in~\eqref{eq:bg_meas_model} is used to recursively track the background state using an \gls{ekf}. The one-step predictive distribution approximated using a Gaussian distribution is given as
\begin{equation}
p(\theta_{j,k}\mid y_{j,1:k-1};\mathcal{H}_0)\approx\mathcal{N}\left(\theta_{j,k};\hat{\theta}_{j,k\mid k-1}, P_{j,k\mid k-1}\right),
\label{eq:bg_predictive_dist}
\end{equation}
where, $\hat{\theta}_{j,k\mid k-1}$ and $P_{j,k\mid k-1}$ are the predicted background-state mean and covariance, respectively, obtained from the \gls{ekf}.

The statistical behavior of the background model is governed by the hyperparameters $\Theta_j=\{\sigma_{w,j},\sigma_{c,j},\sigma_{d,j}\}$, which can be estimated from target-absent measurements using Type-II maximum likelihood estimation~\cite{Bishop2006}. A procedure for the hyperparameter learning and \gls{ekf}-based background tracking is provided in~\cite{Koul2026}.

\section{BERNOULLI FILTER-BASED TARGET TRACKING}\label{sec: algorithm}
The predictive background statistics obtained in \Cref{sec: backgrnd_Model} are now used to approximate the likelihoods required by a Bernoulli track-before-detect filter. This section first summarizes the Bernoulli filtering recursion. A \gls{cfar}-based point-detection likelihood is then introduced as a baseline, followed by the proposed approximate raw-measurement likelihood. The TrBD methodology is presented next, following the presentation in~\cite{Ristic2013}.
\subsection{Bernoulli Filter Overview}
Let $q_k$ denote the probability of existence, and $z_k$ denote the state of the target, given that the target exists. The target state and the probability of its existence can be simultaneously inferred by modeling the target using a Bernoulli random finite set (BRFS)~\cite{Ristic2013}. The BRFS $Z_k$ jointly describes these components using the finite set statistics (FISST) probability density function (PDF)~\cite{Mahler2007, Ristic2013}
\begin{equation}
    f_k(Z)=\begin{cases}
        1-q_k, & \text{if}\, Z=\emptyset\\
        q_k s_k(z), & \text{if}\, Z=\{z\}
    \end{cases}
\end{equation}
where $s_k(z)$ is the PDF of target state $z_k$.  The posterior PDF 
\begin{equation}
    f_k(Z\mid y_{1:k'})=\begin{cases}
        1-q_{k \mid k'}, & \text{if}\, Z=\emptyset\\
        q_{k \mid k'}s_{k \mid k'}(z). & \text{if}\, Z=\{z\}
    \end{cases}
\end{equation}
can be calculated using the Bernoulli filter~\cite{Ristic2013}. Here, $q_{k \mid k'}$ and $s_{k \mid k'}$ are the posterior probability of existence and posterior state distribution for time instant $k$ given measurements up to time $k'$. Given $f_{k-1|k-1}$,  the Bernoulli filter recursion for calculating the posterior distribution $f_{k \mid k}$ are given by the prediction and measurement-update steps~\cite{Ristic2013}
\begin{subequations}\label{eq: prior_TrBD}
\begin{align}
q_{k \mid k-1}
&= p_b\bigl(1-q_{k-1|k-1}\bigr) + p_s\,q_{k-1|k-1},\\
\begin{split}
s_{k \mid k-1}(z)
&= \frac{p_b\bigl(1-q_{k-1|k-1}\bigr)\, b_{k \mid k-1}(z)}{q_{k \mid k-1}}\\
+& \frac{p_s\,q_{k-1|k-1}\int \pi_{k \mid k-1}(z \mid z')\, s_{k-1|k-1}(z')\,dz'}{q_{k \mid k-1}},
\end{split}
\end{align}
\end{subequations}
and
\begin{subequations}\label{eq: posterier_TrBD}
    \begin{align}
        q_{k \mid k}&=\frac{q_{k \mid k-1}\int L(y_k \mid z)s_{k \mid k-1} (z)dz}{1-q_{k \mid k-1}+q_{k \mid k-1}\int L(y_k \mid z)s_{k \mid k-1}(z)dz},\\
        s_{k \mid k}&=\frac{L(y_k \mid z)s_{k \mid k-1}(z)}{\int L(y_k \mid z)s_{k \mid k-1}(z)dz},
    \end{align}
\end{subequations}
respectively. Here, $b_{k \mid k-1}$ is the birth density of the newly born target, and $p_s$ and $p_b$ denote the probabilities of target survival and birth between time steps, respectively. Further, the transition density $\pi_{k \mid k-1}(z \mid z')$ is the PDF of the state $z$ conditioned on $ z'$, describing the target motion, and $L(y_k \mid z)$ is the conditional likelihood ratio given the target state $z$.
 
\subsection{Likelihood with CFAR-Based Detections}\label{subsec: CFAR}
Traditionally, target tracking is performed using point detections obtained from a detector. In active sonar, such detections can be generated by applying a \gls{cfar} detector to a delay--Doppler map obtained from a matched-filter bank. For receiver $j$, the matched-filter output at ping $k$ is 
\begin{equation}
   \chi_j(\tau,\beta,k)  = \lvert\bar{s}^\top(\tau,\beta) \, y_{j,k}\rvert^2,
\end{equation}
where $y_{j,k}$ is the received measurement vector and $\bar{s}(\tau,\beta)$ is the unit-norm delayed and Doppler-scaled replica of the transmitted waveform.

To compensate for the background variations, a cell-averaging \gls{cfar} processor~\cite{Richards2022} is applied to the matched-filter output $\chi_j(\tau,\beta,k)$. Each grid cell is treated as a cell under test (CUT). The training cells surrounding each CUT are used to estimate the local background levels, while guard cells are excluded to limit contamination from a possible target return. Since the background exhibits temporal persistence across consecutive pings, a three-dimensional \gls{cfar} processor is used, with the training and guard regions extending over delay, Doppler scale, and previous pings. A detection is declared when the CUT statistic exceeds the threshold associated with the prescribed false-alarm probability. The estimation of the background level can be viewed as a local background-learning process.

After \gls{cfar} thresholding and peak merging, the detector output at ping $k$ and receiver $j$ is represented as the finite set
\begin{subequations}
\begin{equation}
    \mathcal Y_{j,k} = \left\{\tilde{y}_{j,k,1},\ldots,\tilde{y}_{j,k,M_{j,k}}\right\}
    \end{equation}
    \begin{equation}
    \tilde{y}_{j,k,m} = \begin{bmatrix}
    \tau_{j,k,m}\\
    \beta_{j,k,m}
    \end{bmatrix}
\end{equation}
\end{subequations}
where $\mathcal{M}_{j,k}$ is the number of \gls{cfar} detections. The set $\mathcal Y_{j,k}$ may contain both target-generated detections and false detections.

The false detections are typically modeled as a Poisson point process over the delay--Doppler scale surveillance region $\mathcal D_j$. Let $\lambda_j$ denote the expected number of false detections per ping at receiver $j$, and $\kappa_j(\tilde y)$ denote the spatial probability density of a false detection. Under the target-absent hypothesis, the finite-set likelihood becomes
\begin{equation}\label{eq: l0_CFAR}
    \ell_0(\mathcal Y_{j,k}) = e^{-\lambda_j} \prod_{\tilde y\in\mathcal Y_{j,k}}  \lambda_j\kappa_j(\tilde y).
\end{equation}
When a target with state $z_k$ is present, it is assumed to generate at most one detection at receiver $j$ with probability $p_{d}$. Hence, the likelihood function of measurement $\tilde y$ due to state $z_k$ is given by
\begin{equation}
    g_j(\tilde y|z_k) = \mathcal N\!\left(\tilde{y}; h_j(z_k),R_{c}\right),
\end{equation}
where
\begin{equation}
    h_j(z_k)  =  \begin{bmatrix}  
    \tau_j(z_k)\\ 
    \beta_j(z_k)
    \end{bmatrix}
\end{equation}
is the delay--Doppler scale measurement predicted by the target state, and
\begin{equation}
    R_{c}  =  \begin{bmatrix}
    \sigma_{\tau}^2 & 0\\
    0 & \sigma_{\beta}^2
    \end{bmatrix}
\end{equation}
is the covariance of the \gls{cfar} point-detection error. The delay and Doppler-scale uncertainties are chosen based on the corresponding resolution limits,
\begin{equation}
    \sigma_{\tau} = \frac{1}{BW},  \qquad \sigma_{\beta} = \frac{1}{BW\,T_p},
\end{equation}
where $T_p$ is the transmitted waveform pulse duration. 

Using the standard single-target finite-set measurement model~\cite{Ristic2013}, the likelihood for the measurements when the target is present is given by
\begin{equation}
\begin{aligned}
    \ell_1(\mathcal Y_{j,k}\mid z_k) &=(1-p_{d})\ell_0(\mathcal Y_{j,k})  \\
    &\quad + p_{d} \sum_{\tilde y\in\mathcal Y_{j,k}}  g_j(\tilde y\mid z_k) \ell_0(\mathcal Y_{j,k}\setminus\{\tilde y\}),
\end{aligned}
\end{equation}
where $\setminus$ denotes set difference. Dividing by $\ell_0(\mathcal Y_{j,k})$ gives the receiver-wise likelihood ratio as
\begin{equation}
    L(\mathcal Y_{j,k}\mid z_k) = 1-p_{d} + p_{d} \sum_{\tilde y\in\mathcal Y_{j,k}} \frac{g_j(\tilde y\mid z_k)}{\lambda_j\kappa_j(\tilde y)} .
\end{equation}
Assuming that the receiver-wise detections are conditionally independent given the target state, the multi-receiver likelihood ratio factorizes as
\begin{equation}\label{eq: mult_rec_like}
   \mathcal L(\mathcal Y_{1,k},\ldots,\mathcal Y_{N_r,k}\mid z_k) = \prod_{j=1}^{N_r} L(\mathcal Y_{j,k}\mid z_k),
\end{equation}
where $N_r$ is the number of receivers.
This likelihood ratio can then be used in the Bernoulli filter update~\eqref{eq: posterier_TrBD} to construct a point-detection-based tracker. This point-detection tracker is less computationally demanding than a raw sensor measurement-based tracker as it processes only the finite set of \gls{cfar} detections. However, when a weak target echo is embedded in the time-varying background, \gls{cfar} may miss the target or generate false detections from the background. Missed target detections can reduce the posterior existence probability, whereas persistent background-generated detections may initiate false tracks or bias the inferred target state.

\subsection{Likelihood for the Proposed Method}\label{sec:background_aware_tracking}
Using the predicted background distribution in~\eqref{eq:bg_predictive_dist} together with the measurement model in~\eqref{eq:reparam_stat_model_clean}, the likelihood under the target present hypothesis can be approximated as
\begin{equation}\label{eq:target_present_marginal_likelihood}
\begin{split}
\ell_1(y_{j,k}\mid z_k,y_{j,1:k-1}) \approx & \int p(y_{j,k}\mid z_k,\theta_{j,k})\\
 & \times p(\theta_{j,k}\mid y_{j,1:k-1};\mathcal H_0) \,d\theta_{j,k}.    
\end{split}
\end{equation}
Here, the background predictive distribution is evaluated under $\mathcal{H}_0$ to separate background inference from target inference. 

Since the measurement covariance $R(\theta_{j,k})$ depends on the unknown background state vector, the integral in~\eqref{eq:target_present_marginal_likelihood} does not generally admit a closed-form solution. However, if the covariance is evaluated at the predicted background mean, i.e.,
\begin{equation}\label{eq:covariance_mean_approximation}
R(\theta_{j,k})\approx R(\hat{\theta}_{j,k\mid k-1}),
\end{equation}
the integral can be evaluated analytically. This results in the approximate likelihood
\begin{equation}\label{eq:target_present_likelihood}
\begin{split}
 \ell_1(y_{j,k}\mid z_k,y_{j,1:k-1})\approx\mathcal{N}\!\Big(y_{j,k}; & \, H\hat{\theta}_{j,k\mid k-1} \\
             & + x_j^o(z_k), \Sigma_{j,k}\Big).   
\end{split}
\end{equation}
The corresponding target-absent likelihood is obtained by setting $x_j^o(z_k)=0$ as
\begin{equation}\label{eq:target_absent_likelihood}
\ell_0(y_{j,k} \mid y_{j,1:k-1})\approx\mathcal{N}\!\left(y_{j,k};H\hat{\theta}_{j,k\mid k-1},\Sigma_{j,k}\right),
\end{equation}
where $$\Sigma_{j,k}= HP_{j,k\mid k-1}H^\top+ R(\hat{\theta}_{j,k \mid k-1}).$$
Further, the corresponding receiver-wise likelihood ratio can be written as
\begin{equation}\label{eq:bg_aware_like_ratio}
\begin{split}
L(y_{j,k}\mid y_{j,1:k-1},z_k)=\exp\!\Big(&{x_j^o}^{\top}(z_k)\Sigma_{j,k}^{-1}\nu_{j,k}\\
&-\frac{1}{2}{x_j^o}^{\top}(z_k)\Sigma_{j,k}^{-1}x_j^o(z_k)\Big),
\end{split}
\end{equation}
where $ \nu_{j,k}=y_{j,k}-H \hat {\theta}_{j, k\mid k-1}$ is the background-compensated measurement. For multiple receivers, the receiver-wise likelihood ratios are combined according to~\eqref{eq: mult_rec_like}.

The proposed method for the approximation of the measurement likelihood leads to a computationally efficient inference scheme. The background is inferred separately under the target-free assumption, and the background predicted statistics are then used to define the measurement likelihoods. This reduces the computational burden for the \gls{trbd} filter as compared to joint inference over both the target and background state, at the cost of neglecting their posterior dependence.

The same EKF inferred background statistics can also be used to construct a background-compensated \gls{cfar} detector. In that case, the matched-filter input is replaced by the background-compensated measurement vector $\nu_{j,k}$ and the delay--Doppler statistic becomes
\begin{equation}
\chi_j^{\mathrm{BC}}(\tau,\beta,k)=\left \lvert \bar{s}^\top(\tau,\beta)\nu_{j,k}\right \rvert ^2 .
\end{equation}
This modified detector is denoted by \gls{cfar}-BC in the later presented performance evaluation.

\section{TARGET STATE DYNAMICS AND BIRTH MODEL}
 In the Bernoulli TrBD framework, both the target state transition density $\pi(z \mid z')$ and the birth density $b_{k \mid k-1}(z)$ are required. These densities are defined as follows.

 \subsection{Motion Model}
The multistatic sensor geometry enables the target to be represented and tracked in Cartesian coordinates. To do so, let the target state at the $k$:th ping be 
\begin{equation}\label{eq: target_states}
z_k=\begin{bmatrix}
    \vec{p}_k^\top & \vec{v}_k^\top & \eta_k^{\text{\footnotesize(dB)}} 
\end{bmatrix}^\top,    
\end{equation}
where $\vec{p}_k$ and $\vec{v}_k$ are the target position and velocity, respectively. Further, $\eta_k^{\text{\footnotesize(dB)}}$ is the reflected target power (in dB) measured 1 m away from the target. 

In this work, the target motion is modeled using a constant-velocity model, whereas the reflected target power follows a random walk model. The resulting transition density is
\begin{equation}
    \pi(z_{k+1}| z_{k}) = \mathcal{N}(z_{k+1}; Fz_k, GQ_xG^T),
\end{equation}
where
\begin{equation}
    F =
    \begin{bmatrix}
        I & \Delta_t I & 0\\
        0 & I & 0\\
        0 & 0 & 1
    \end{bmatrix},
    \qquad
    G =
    \begin{bmatrix}
        \frac{\Delta_t^2}{2} I & 0\\
        \Delta_t I & 0\\
        0 & 1
    \end{bmatrix},
\end{equation}
and
\begin{equation}
    Q_x =
    \begin{bmatrix}
        \sigma_a^2 I & 0 \\
        0 & \sigma^2_{\eta^{\text{\footnotesize(dB)}}}
    \end{bmatrix}.
\end{equation}
Here, $\Delta_t$ is the time duration between the consecutive pings. Further, $\sigma^2_{a}$ and $\sigma^2_{\eta^{\text{\footnotesize(dB)}}}$ are the process noise variances for the acceleration component and the reflected target power, respectively.

Based on the target state $z$, the target signal parameters $\tau_{o,j}$, $\beta_{o,j}$, and $a_{o,j}$ can be computed using
\begin{subequations}
\begin{align}
    \tau_{o,j}(z) &= \frac{1}{c_w} \left( \|\vec{p}-\vec{p}_{T}\| + \|\vec{p}-\vec{p}_{R_j}\| \right), \\
    \beta_{o,j}(z) &= 1 - \frac{1}{c_w} \left( \frac{\vec{p}-\vec{p}_{T}} {\|\vec{p}-\vec{p}_{T}\|} + \frac{\vec{p}-\vec{p}_{R_j}} { \|\vec{p}-\vec{p}_{R_j}\|} \right)^\top \vec{v}, \\
   a_{o,j}(z)& = \sqrt{10^{(\eta^{\text{\footnotesize(dB)}} - \mathrm{TL}_{j})/10}},
\end{align}
\end{subequations}
where $c_w$ is the speed of sound in water, and $\vec{p}_T$ and $\vec{p}_{R_j}$ are the positions of the transmitter, and the $j$:th receiver, respectively. The reflected target power $\eta^{\text{\footnotesize(dB)}}$ is defined based on the sonar link budget equation as
\begin{equation}
    \eta^{\text{\footnotesize(dB)}} = \mathrm{SL} -\mathrm{TL}_T+\mathrm{TS}
\end{equation}
where $\mathrm{SL}$ is the transmitted signal power level, and $\mathrm{TS}$ is the target strength. Further, $\mathrm{TL}_T$ and $\mathrm{TL}_j$ denote the transmission losses from the transmitter to the target and from the target to the $j$:th receiver, respectively, and are assumed constant.
\begin{figure*}[t]
    \centering
    \includegraphics[width=1.04\textwidth,height=0.28\textwidth]{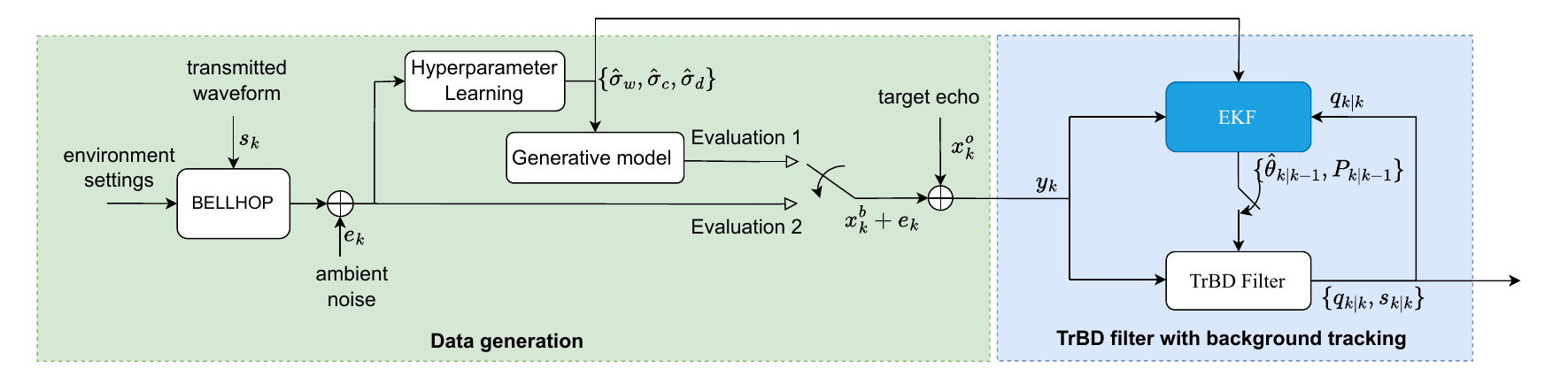}
    \caption{Workflow for evaluating the proposed background-tracking and \gls{trbd} framework. The proposed method is evaluated using two sets of measurements. The measurements from the generative model are conditioned on the hyperparameters learned from the target-free BELLHOP-generated measurements. Both measurement sets are processed by the \gls{ekf} background tracker and the Bernoulli \gls{trbd} filter.}
    \label{fig:flow_chart}
\end{figure*}

\subsection{Birth Model}
The birth model describes the state distribution of a newly appearing target. In the Bernoulli filter, the birth density should place new particles in regions of the state space that are physically plausible and consistent with the received measurements. Since \( y_{k-1}\) provides information about the target position, velocity, and target-return level through the corresponding delay, Doppler scale, and amplitude, respectively, the birth density is modeled as
\begin{equation}
    b_{k \mid k-1}(z) \propto L(y_{k-1}\mid z)\, \pi_B(z),
\end{equation}
where \(L(y_{k-1}\mid z)\) is the likelihood ratio evaluated for the candidate target state $z$, and $\pi_B(z)$ is the prior birth density. The prior birth density can be factorized as
\begin{equation}
    \pi_B(z) = f_B(\vec{v} \mid \vec{p})\, f_B(\vec{p})\,f_B\!\left(\eta^{(\mathrm{dB})}\right).
\end{equation}
The target position $\vec{p}$ is assumed to be uniformly
distributed over the surveillance region and the reflected target power
is also assigned a uniform prior in dB as
\begin{equation}
    f_B\!\left(\eta^{(\mathrm{dB})}\right) =\mathcal U \left(\eta^{(\mathrm{dB})};\eta_{\min},\eta_{\max}\right).
\end{equation}
In the considered reverberation-limited shallow-water environment,
the interval \([\eta_{\min},\eta_{\max}]\) is selected to cover the
expected range of weak target returns.

The velocity prior is conditioned on the target position. The speed of a newly appearing target is assumed to be uniformly distributed over an admissible speed interval. At the same time, its direction is selected based on the target's position relative to the multistatic sensor geometry. In particular, newly appearing targets are assumed to move approximately toward the multistatic sensor configuration.

\section{EVALUATION}\label{sec:eval}
To evaluate the proposed likelihood formulation, two sets of raw sensor measurements are considered. The first set is generated from the statistical model described in~\eqref{eq:reparam_stat_model_clean} and~\eqref{eq:back_dyn_model} and is used to assess tracking performance under matched-background model conditions, since the same model is used to derive the approximate likelihood employed by the TrBD filter.

The second set is constructed using time-varying multipath propagation data generated with BELLHOP~\cite{porter2011bellhop}. The BELLHOP-generated multipath data is combined with the target return and ambient noise to form the raw sensor measurements. Since the multipath propagation data is not generated from the assumed statistical background model, the resulting measurements are used to assess the robustness of the proposed tracking methodology under background-model mismatch.

The objective of these evaluations is to assess the effectiveness of the proposed approximate likelihood for target tracking under both matched-background and background-model-mismatch conditions, rather than to quantify the approximation error associated with each modeling assumption separately. The overall process of generating the raw sensor measurements and evaluating the proposed likelihood approximation within the Bernoulli TrBD framework is illustrated in Fig.~\ref{fig:flow_chart}.

To investigate the effect of possible target energy absorption in the background inference process, three background-update strategies are considered. In the first strategy, the \gls{ekf} measurement update is performed at every ping, allowing the tracked background statistics to adapt continuously to the time-varying background. In the second strategy, the measurement update is stopped at the ping at which the target appears, and the background state and its uncertainties are subsequently propagated using only the prediction step. Since the target-arrival ping is generally unknown in practice, a third strategy is employed in which the \gls{ekf} measurement update is skipped at ping $k$ whenever $q_{k\mid k}\geq\gamma$, where $\gamma$ is the track-confirmation threshold. If $q_{k\mid k}$ subsequently falls below $\gamma$, the measurement update is resumed. This provides a practical mechanism for preventing the \gls{ekf} from adapting to the target echo while the target-existence probability remains above the confirmation threshold.

For comparison, two \gls{cfar}-based trackers are considered. The conventional \gls{cfar}-based tracker described in~\Cref{subsec: CFAR} and the background-compensated \gls{cfar}-BC tracker described in~\Cref{sec:background_aware_tracking}. For the \gls{cfar}-BC tracker, the compensation is performed based on the background-update strategies described earlier. The same measurement datasets and Bernoulli filtering framework are used for all tracking methods.

\subsection{Performance Evaluation Metrics}
The \gls{trbd} filter recursively infers the posterior target-existence probability $q_{k\mid k}$ and the target-state posterior distribution, from which the minimum mean-square estimate $\hat{z}_{k\mid k}$ is calculated. Therefore, the performance metrics should account for both target confirmation and localization accuracy. The localization accuracy is evaluated using the \gls{gospa} metric~\cite{Rahmathullah2017}, whereas the target-confirmation performance is evaluated using the probability of track confirmation ($P_{\mathrm{TC}}$), probability of false track confirmation ($P_{\mathrm{FTC}}$), and mean time to track confirmation ($\mathrm{MTTC}$). These quantities are defined as follows.
 
\subsubsection{Single-target GOSPA metric}
To jointly penalize localization error, missed target confirmations, and false target confirmations, we use a \gls{gospa} metric computed from the true and estimated sets $Z_k$ and $\bar{Z}_{k \mid k}$. Since the considered problem is a single-target problem, both sets can have cardinality zero or one. For the case of a single target and  $|\bar Z_{k \mid k}| \le \lvert Z_k \rvert$, the \gls{gospa} metric can be written with a slight abuse of notation as
\begin{equation}
\scalebox{0.85}{$\displaystyle
\mathfrak{d}_2^{(\mathfrak{c},\alpha)}
(\bar{Z}_{k \mid k},Z_k)= \left(\sum_{\hat{z}\in\bar{Z}_{k \mid k}} \mathfrak{d}^{(\mathfrak{c})}(\hat{z},z_k)^2 + \frac{\mathfrak{c}^2}{\alpha}\left(\lvert Z_k \rvert-\lvert\bar{Z}_{k \mid k}\rvert\right) \right)^{1/2}$}
\end{equation}
 where $z_k$ is the ground truth target state, $\mathfrak{d}^{(\mathfrak{c})}(\hat{z},z_k)= \min (||\hat{\vec{p}}-\vec{p}_k||_2,\mathfrak{c})$ is the distance measure with cut-off $\mathfrak{c}$. 
Here, $\bar{Z}_{k \mid k}$ is the set of confirmed targets 
\begin{equation}\label{eq: confirmation_target}
    \bar{Z}_{k \mid k}=\begin{cases}
\{\hat z_{k \mid k}\}, & q_{k \mid k}\geq \gamma,\\
\emptyset, & q_{k \mid k}< \gamma.
\end{cases}
\end{equation}
where $\gamma=0.96$, and $\mathfrak{c}=150$ m and $\alpha=2$. If $|\bar{Z}_{k \mid k}|>\lvert Z_k \rvert$, $\mathfrak{d}_2^{(\mathfrak{c},\alpha)}(\bar{Z}_{k \mid k},Z_k) \triangleq \mathfrak{d}_2^{(\mathfrak{c},\alpha)}(Z_k,\bar{Z}_{k \mid k})$. Here, the ground-truth target set is
\begin{equation}
    {Z}_{k}=\begin{cases}
\emptyset, & k < \bar k, \\
\{z_{k}\}, & k \geq \bar k,
\end{cases}
\end{equation}
and $\bar k$ is the ping index at which the target appears in the measurement.

\subsubsection{Probability of track confirmation and false track confirmation}
The $\mathrm{P}_{\mathrm{TC}}$ characterizes the probability of confirming a track under the target-present hypothesis, whereas $\mathrm{P}_{\mathrm{FTC}}$ characterizes the probability of confirming a track under the target-absent hypothesis. These quantities are defined as
\begin{subequations}\label{eq:ptc_pftc}
\begin{align}
P_{\mathrm{TC}}&=\frac{1}{N_{\mathrm{MC}}}\sum_{i=1}^{N_{\mathrm{MC}}}\mathbb{I}\left\{\max_{\bar k\leq k\leq N_k}q_{k\mid k}^{(i)}\geq\gamma\,\middle|\,\mathcal{H}_1\right\},
\label{eq:ptc}\\
P_{\mathrm{FTC}}&=\frac{1}{N_{\mathrm{MC}}}\sum_{i=1}^{N_{\mathrm{MC}}}\mathbb{I}\left\{\max_{1\leq k\leq N_k}q_{k\mid k}^{(i)}\geq\gamma\,\middle|\,\mathcal{H}_0\right\}.
\label{eq:pftc}
\end{align}
\end{subequations}
Here $\mathbb{I}\{\cdot\}$ is the indicator function, $N_{\mathrm{MC}}$ is the number of Monte Carlo realizations, and $N_k$ is the total number of pings. 

\subsubsection{Mean time-to-track-confirmation}
The $\mathrm{MTTC}$ characterizes the average ping delay between target appearance and the first track confirmation under $\mathcal{H}_1$. Since a track may not be confirmed in every realization, the $\mathrm{MTTC}$ is computed only over the realizations in which track confirmation occurs. To do so,  let $N_{\mathrm{conf}}$ denote the number of such realizations. The $\mathrm{MTTC}$ is then computed  as
\begin{equation}\label{eq:mttc}
\mathrm{MTTC}=\frac{1}{N_{\mathrm{conf}}}\sum_{i=1}^{N_{\mathrm{conf}}}\left(\min\left\{k\geq\bar{k}:q_{k\mid k}^{(i)}\geq\gamma\right\}-\bar{k}\right).
\end{equation}

\subsection{Simulation Setup}

\begin{figure}[t]
    \centering
    \includegraphics[width=0.46\textwidth, height=0.4\textwidth]{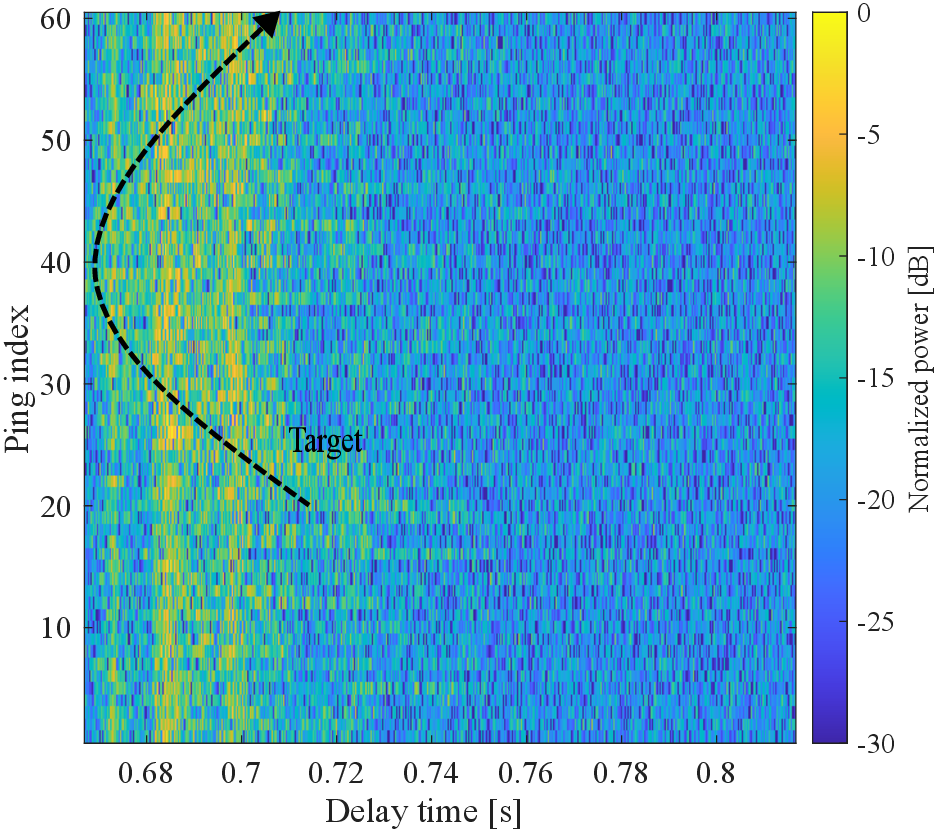}
    \caption{Simulated raw sensor measurements using the BELLHOP-generated multipath propagation for receiver~$1$. The dashed curve indicates the true target trajectory with the target appearing at ping index 20.}
    \label{fig:measurements}
\end{figure}

In the current work, one transmitter and two receivers are considered. The target moves at a constant speed of $5$~[m/s] along a trajectory that crosses the direct transmitter--receiver-1 path. This trajectory is selected to evaluate the tracking methods when the target echo is masked by the structured and time-varying multipath component of the background. The target is introduced after an initial target-free interval, allowing false track confirmations to be evaluated before target appearance and the track-confirmation delay to be measured afterward. To illustrate the considered scenario, Fig.~\ref{fig:measurements} shows the raw sensor measurements at receiver~1. The measurements contain the target return, BELLHOP-generated time-varying multipath signal propagation, and ambient noise. The environmental configuration and signal parameters used for the BELLHOP simulations are summarized in Table~\ref{tab:bellhop_config}.
\begin{table}[t]
\centering
\caption{Environment configuration and system properties for the generation of measurements in the multipath channel.}
\label{tab:bellhop_config}
\footnotesize
\renewcommand{\arraystretch}{1.05}
\resizebox{\columnwidth}{!}{%
\begin{tabular}{@{} l l @{}}
\toprule
\textbf{Parameter} & \textbf{Value} \\
\midrule
Signal type & Linear frequency-modulated (LFM) chirp \\
Number of pings $N_k$ & $60$ \\
Pulse duration $T_p$ & $30$ ms \\
Pulse repetition interval (PRI) & $1.4$ s \\
Carrier frequency $f_c$ & $3$ kHz \\
Sampling frequency $1/T_s$ & $15$ kHz \\
Bandwidth (BW) & $4$ kHz \\
Ocean depth & $50$ m \\
Field generation mode & Time varying \\
Sound speed profile & Iso-velocity ($1500$ m/s) \\
Bottom type & Acoustic half-space \\
Bottom density & $1.7$ g/cm$^{3}$ \\
Bottom attenuation & $0.5$ dB/$\lambda$ \\
Bottom roughness & $0.05$ \\
Sound speed in bottom & $1650$ m/s \\
Distance between sensor nodes & $1$ km \\
Sensor node depth & $4$ m from ocean surface \\
Sea-surface spectrum & JONSWAP~\cite{hasselmann1973measurements} \\
Significant wave height $H_s$ & $2$ m \\
Transmitter position & $[0, 0]^\top$ m \\
Receiver 1 position  & $[1000, 0]^\top$ m \\
Receiver 2 position  & $[500, -866.7]^\top$ m \\
\bottomrule
\end{tabular}}
\end{table}

\subsubsection{Signal-to-noise-ratio}
A conventional \gls{snr} does not adequately characterize the operating conditions for the considered tracking problem because the background is colored, time-varying, and different across receivers. Therefore, an alternative \gls{snr} definition is required to account for both the target-signal energy and the predicted measurement covariance across all receivers. To do so, the effective \gls{snr} is defined as
\begin{equation}\label{eq:SNR_eff}
\mathrm{SNR}_{\mathrm{eff}} \triangleq 10\log_{10}\left(\sum_{k=\bar{k}}^{N_k}\sum_{j=1}^{N_r}{\bar{x}_j^{o}(z_k)}^\top\Sigma_{j,k}^{-1}\bar{x}_j^{o}(z_k)\right),
\end{equation}
where $\bar{x}_j^{o}(z_k)=a_{o,j}\bar{s}(\tau_{o,j},\beta_{o,j})$, $\bar{s}(\cdot)$ is the normalized target-signal replica, $\bar{k}$ is the ping at which the target appears, $N_k$ is the total number of pings, and $N_r$ is the number of receivers. The resulting \gls{snr}$_{\mathrm{eff}}$ quantifies the cumulative target-signal energy relative to the time-varying measurement covariance, accumulated across all receivers.

\subsection{Parameter Settings}
The Bernoulli \gls{trbd} filter is implemented using a particle filter. The filter uses a small birth probability and a high survival probability to represent rare target appearance and persistent target motion once the target is present. To generate the point detections using \gls{cfar}, the training and guard regions are selected to suppress detections arising from local multipath structure. Closely spaced detections are merged using a neighborhood selected according to the delay and Doppler-scale resolutions. The common parameter settings for the \gls{trbd} filter used by all the trackers and the parameters specific to the \gls{cfar} detector are summarized in Table~\ref{tab:tracking_config}. Here, the parameters $\lambda_j$ and $p_d$ used in the \gls{cfar}-based likelihood were selected empirically through preliminary experiments and were subsequently kept fixed throughout the evaluations. The spatial density of false detections, $\kappa_j(\cdot)$, is assumed to be uniform over the delay--Doppler-scale surveillance region. Furthermore, the track-confirmation threshold $\gamma$ is also kept fixed for all tracking methods so that the same posterior-existence decision criterion is applied throughout the evaluations.
\begin{table}[t]
\centering
\caption{Common \gls{trbd} filter parameters and additional parameters used by the \gls{cfar}-based trackers.}
\label{tab:tracking_config}
\renewcommand{\arraystretch}{1}
\begin{tabularx}{\columnwidth}{@{} 
>{\centering\arraybackslash}p{0.18\columnwidth}
>{\centering\arraybackslash}p{0.25\columnwidth}
X
@{}}
\toprule
\textbf{Sym.} & \textbf{Value} & \textbf{Description} \\
\midrule

\multicolumn{3}{c}{\textit{(a) Common TrBD filter parameters}} \\
\midrule
$p_b$ & $10^{-3}$ & Probability of target birth. \\
$p_s$ & $1-10^{-12}$ & Probability of target survival. \\
$N_S$ & $1.5 \times 10^4$ & Number of surviving particles. \\
$N_B$ & $1.5 \times 10^4$ & Number of birth particles. \\
$\sigma_{a}$ & 0.1 & Process-noise standard deviation. \\
$\sigma_{\eta}^{\textrm{(dB)}}$ & 1 & Process-noise standard deviation for reflected target power $\eta^{\textrm{(dB)}}$. \\
$\gamma$ & 0.96 & Target confirmation threshold. \\
$[\eta_{\text{min}}, \eta_{\text{max}}]$ & [-25, 5]  & Lower and upper limits, in dB, for the reflected target power uniform prior.\\

\midrule
\multicolumn{3}{c}{\textit{(b) \gls{cfar}-detector and likelihood parameters}} \\
\midrule
\multicolumn{2}{c}{$[2,\,42,\,40]$} & Guard cells in the delay, Doppler-scale, and temporal dimensions. \\

\multicolumn{2}{c}{$[2,\,126,\,40]$} & Training cells in the delay, Doppler-scale, and temporal dimensions. \\

\multicolumn{2}{c}{$10^{-3}$} & \gls{cfar} false-alarm probability per CUT. \\

\multicolumn{2}{c}{10} & Mean number of false detections per ping, $\lambda_j$. \\

\multicolumn{2}{c}{0.9} & Target-detection probability, $p_d$. \\

\bottomrule
\end{tabularx}
\end{table}
  
\section{RESULTS AND DISCUSSION}
This section presents the results obtained using the two measurement sets described in \Cref{sec:eval}. Evaluation~1 assesses the tracking performance of the proposed likelihood formulation under matched-background model conditions, whereas Evaluation~2 assesses its robustness to mismatch between the assumed statistical background model and the BELLHOP-generated time-varying multipath propagation. All results for Evaluation~1 are obtained using $N_{\mathrm{MC}}=128$ Monte Carlo realizations to characterize the tracking performance under matched-background model conditions. Evaluation~2 is performed using five independently generated time-varying multipath signal propagation scenarios in BELLHOP to assess the robustness of the proposed methodology under background-model mismatch.

\subsection{Evaluation 1}\label{subsec: eval1}
The track-confirmation threshold was selected by examining the $P_{\mathrm{TC}}$--$P_{\mathrm{FTC}}$ curves over a range of $\gamma$. A value of $\gamma=0.96$ was chosen as a conservative operating point and subsequently kept fixed for all methods. Since the same Bernoulli filtering recursion is used throughout, this provides a common track-confirmation decision criterion and avoids method-specific tuning of the posterior-existence threshold.

\begin{figure}[t]
    \centering
    \includegraphics[width=0.5\textwidth, height=0.46\textwidth]{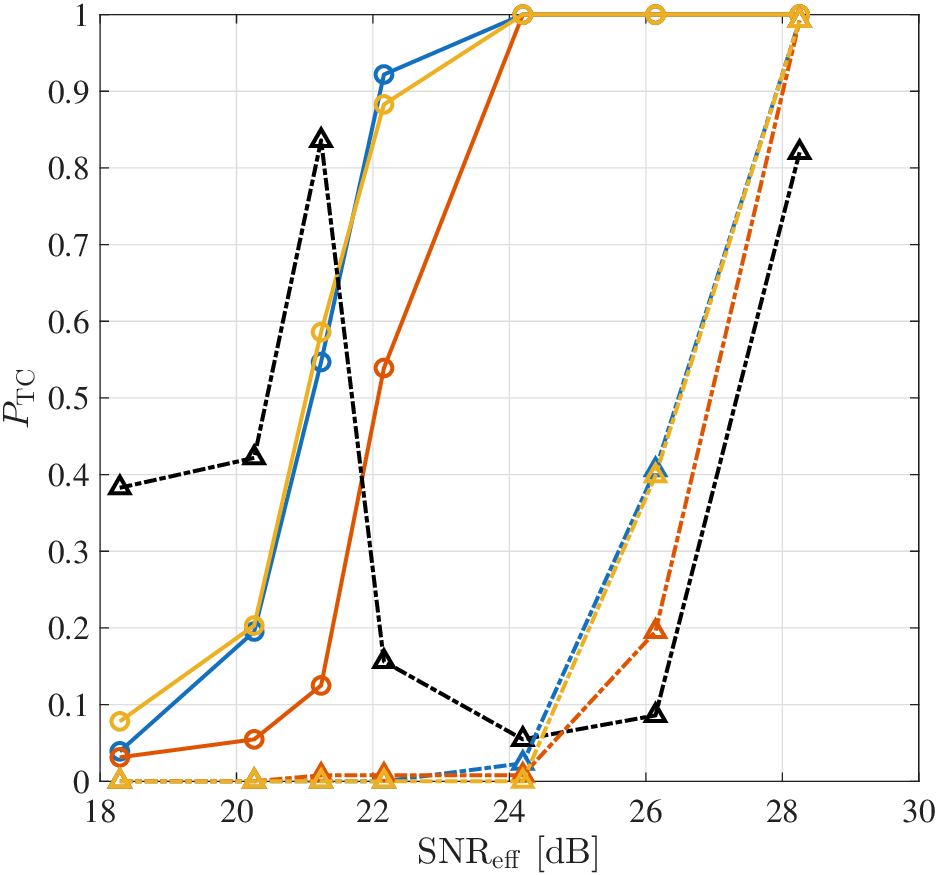}
    \caption{Track-confirmation probability \(P_{\mathrm{TC}}\) vs. \gls{snr}$_{\mathrm{eff}}$.}
    \label{fig:ptc-vs-snreff-all-methods}
\end{figure}

\begin{figure}[t]
    \centering
    \includegraphics[width=0.5\textwidth, height=0.45\textwidth]{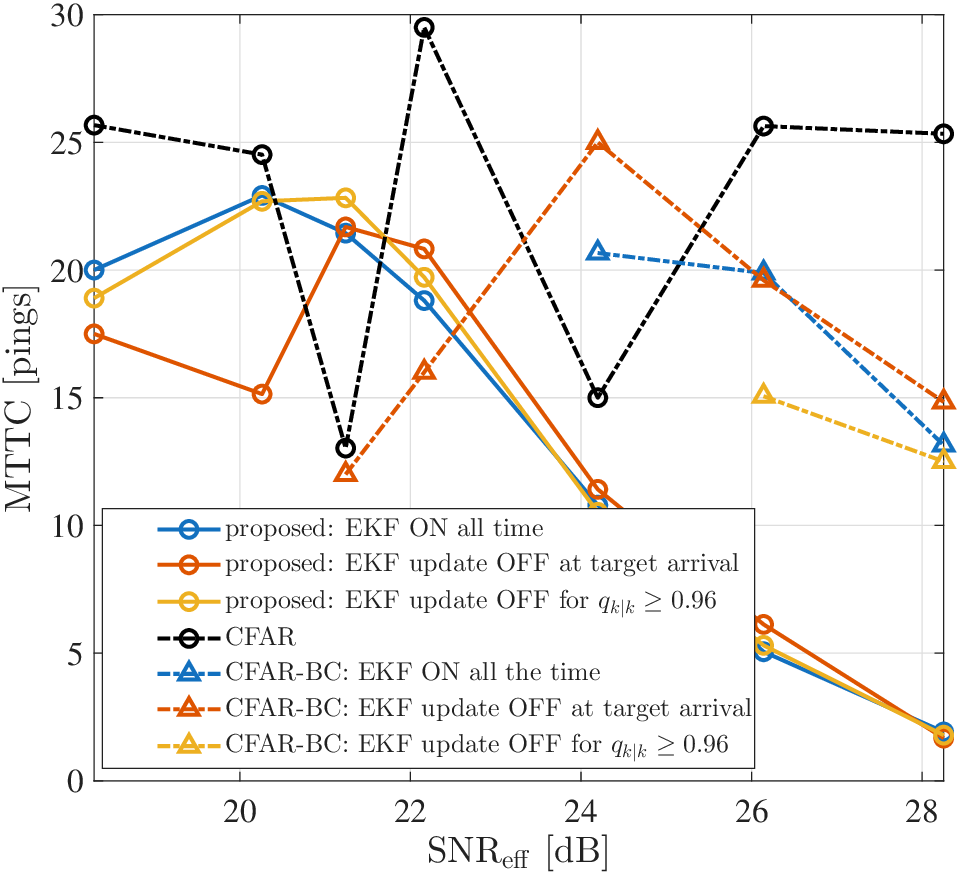}
    \caption{Conditional $\mathrm{MTTC}$ vs. \gls{snr}$_{\mathrm{eff}}$.}
    \label{fig:mttc-vs-snreff-all-methods}
\end{figure}

Figs.~\ref{fig:ptc-vs-snreff-all-methods} and~\ref{fig:mttc-vs-snreff-all-methods} show $P_{\mathrm{TC}}$ and the conditional $\mathrm{MTTC}$, respectively, as functions of \gls{snr}$_{\mathrm{eff}}$ for the different likelihood formulations and background-update strategies. The methods based on the proposed likelihood formulation achieve higher $P_{\mathrm{TC}}$ at lower \gls{snr}$_{\mathrm{eff}}$ values than the \gls{cfar}-based methods. The benefit of tracking the background statistics is also reflected in the performance of the \gls{cfar}-BC tracker relative to the conventional \gls{cfar} tracker. The proposed methods also generally achieve lower $\mathrm{MTTC}$ than the \gls{cfar}-based methods.

It should be noted that $P_{\mathrm{TC}}$ indicates whether a track is confirmed under the target-present hypothesis, while $\mathrm{MTTC}$ is computed only over realizations in which a track is confirmed. Therefore, neither metric, by itself, indicates whether the confirmed track corresponds to the true target. Consequently, the conventional \gls{cfar}-based tracker may exhibit a relatively high $P_{\mathrm{TC}}$ or an apparently small $\mathrm{MTTC}$ due to tracks initiated by multipath-induced detections. This behavior is partly mitigated by \gls{cfar}-BC, which compensates for the tracked background before forming the point detections. The localization accuracy and quality of the resulting tracks are therefore further assessed using the \gls{gospa} metric.

\begin{figure}[t]
    \centering
    \includegraphics[width=0.45\textwidth, height=0.36\textwidth]{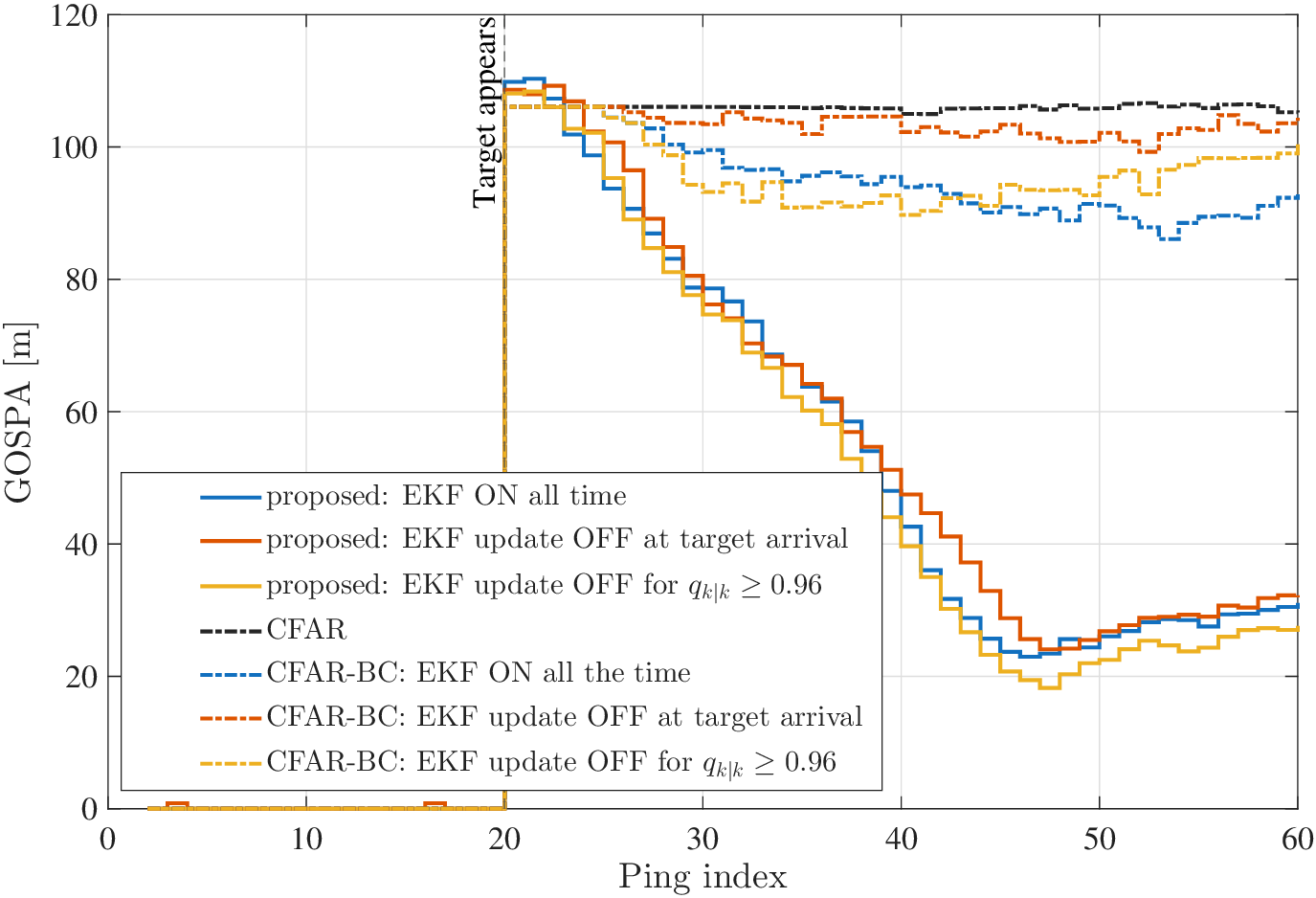}
    \caption{\gls{gospa} vs. ping index at \gls{snr}$_{\mathrm{eff}}=24.3\,[\mathrm{dB}]$.}
    \label{fig:gospa-vs-ping-snreff-24p2}
\end{figure}
\begin{figure}[t]
    \centering
    \includegraphics[width=0.45\textwidth, height=0.4\textwidth]{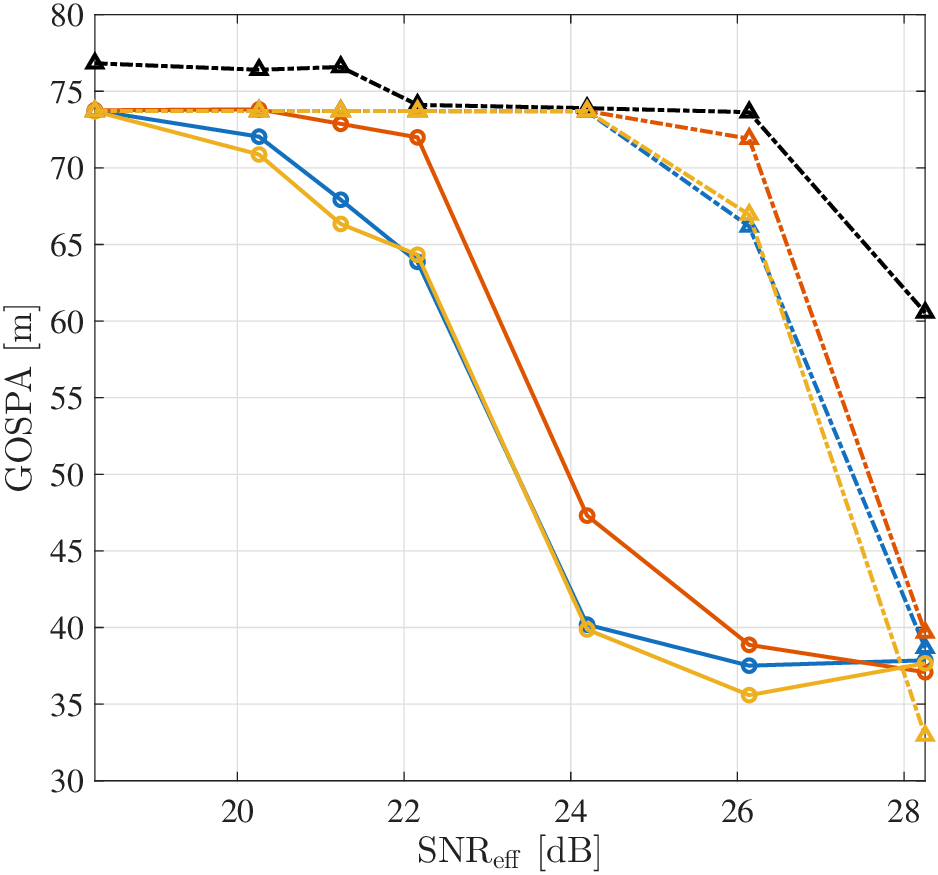}
    \caption{\gls{gospa} versus \gls{snr}$_{\mathrm{eff}}$, averaged over pings and Monte Carlo realizations.}
    \label{fig:gospa-vs-snreff-all-methods}
\end{figure}

Figs.~\ref{fig:gospa-vs-ping-snreff-24p2} and~\ref{fig:gospa-vs-snreff-all-methods} show the \gls{gospa} as a function of ping index at \gls{snr}$_{\mathrm{eff}}=24.3$ [dB] and as a function of \gls{snr}$_{\mathrm{eff}}$, respectively. At \gls{snr}$_{\mathrm{eff}}=24.3$ [dB], the \gls{cfar}-based methods produce large \gls{gospa} values after target appearance. In general, large \gls{gospa} values may result from missed tracks, false tracks, or incorrectly localized target estimates. As shown in Fig.~\ref{fig:gospa-vs-snreff-all-methods}, the methods based on the proposed likelihood formulation achieve lower \gls{gospa} values over most of the considered \gls{snr}$_{\mathrm{eff}}$ range. The values in Fig.~\ref{fig:gospa-vs-snreff-all-methods} are averaged over all pings
and Monte Carlo realizations.

\subsection{Evaluation 2}\label{subsec: eval2}
Table~\ref{tab:bellhop_ping_performance} summarizes the tracking performance of the proposed and \gls{cfar}-based methods for the different background-update strategies using the BELLHOP-generated measurements. The results are reported in terms of the posterior target-existence probability $q_{k\mid k}$ and the \gls{gospa} metric at different ping indices, averaged over all the datasets.

For the proposed likelihood-approximation methods, the posterior target-existence probability generally remains below the confirmation threshold before the target appears at $k=20$, although occasional increases are observed. These transient increases may arise from mismatch between the assumed likelihood model and the time-varying background. After the target appears, the posterior existence probability increases progressively as evidence for target presence accumulates over successive pings. In particular, the strategy in which the \gls{ekf} measurement update is skipped whenever $q_{k\mid k}\geq 0.96$ achieves a high probability of existence more rapidly than the other proposed methods. In contrast, the \gls{cfar}-based methods show low values of target-existence probabilities, with increases occurring only over limited ping intervals.

\begin{table*}[t]
\centering
\caption{Probability of existence and \gls{gospa} as functions of the ping index for the different methods at \gls{snr}$_{\mathrm{eff}}=26.3$ [dB]. The reported values are averaged over five BELLHOP multipath propagation scenarios.}
\label{tab:bellhop_ping_performance}

\small
\setlength{\tabcolsep}{3pt}
\renewcommand{\arraystretch}{1.05}


\begin{tabularx}{0.98\textwidth}{
@{}
L{0.34\textwidth}
!{\vrule width 0.5pt}
*{6}{C}
@{}
}

\hline

\multicolumn{7}{c}{
\textbf{(a) Probability of existence $q_{k|k}$}
} \\

\hline

\diagbox[width=\linewidth]{Method}{Ping index}
& $10$
& $20^{*}$
& $30$
& $40$
& $50$
& $60$
\\

\hline

proposed: EKF ON all time
& \qgtwo{0.200}
& \qgzero{0.025}
& \qgzero{0.004}
& \qsix{0.600}
& \qeight{0.802}
& \qhigh{0.994}
\\

proposed: EKF update OFF at target arrival
& \qgzero{0.076}
& \qgone{0.182}
& \qgtwo{0.237}
& \qsix{0.600}
& \qeight{0.809}
& \qnine{0.937}
\\

proposed: EKF update OFF for $q_{k|k} \geq 0.96$
& \qgtwo{0.200}
& \qgzero{0.024}
& \qgfour{0.437}
& \qhigh{0.996}
& \qhigh{1.000}
& \qhigh{0.971}
\\

\hline

\gls{cfar}
& \qgzero{0.011}
& \qgzero{0.002}
& \qgone{0.107}
& \qgthree{0.307}
& \qsix{0.603}
& \qgfour{0.491}
\\

\gls{cfar}-BC: EKF ON all the time
& \qgzero{0.000}
& \qgzero{0.001}
& \qgthree{0.364}
& \qgzero{0.000}
& \qgzero{0.002}
& \qgtwo{0.200}
\\

\gls{cfar}-BC: EKF update OFF at target arrival
& \qgzero{0.000}
& \qgzero{0.001}
& \qgzero{0.025}
& \qgzero{0.001}
& \qgzero{0.000}
& \qgzero{0.002}
\\

\gls{cfar}-BC: EKF update OFF for $q_{k|k} \geq 0.96$
& \qgzero{0.003}
& \qgzero{0.002}
& \qfive{0.500}
& \qgtwo{0.228}
& \qsix{0.600}
& \qgzero{0.000}
\\

\hline

\end{tabularx}

\vspace{2mm}


\begin{tabularx}{0.98\textwidth}{
@{}
L{0.34\textwidth}
!{\vrule width 0.5pt}
*{6}{C}
@{}
}

\hline

\multicolumn{7}{c}{
\textbf{(b) \gls{gospa} [m]}
} \\

\hline

\diagbox[width=\linewidth]{Method}{Ping index}
& $10$
& $20^{*}$
& $30$
& $40$
& $50$
& $60$
\\

\hline

proposed: EKF ON all time
& \qgtwo{21.21}
& \qgzero{106.06}
& \qgzero{106.06}
& \qsix{59.51}
& \qeight{42.45}
& \qhigh{66.50}
\\

proposed: EKF update OFF at target arrival
& \qgzero{0.00}
& \qgone{106.06}
& \qgtwo{114.85}
& \qsix{76.18}
& \qeight{60.46}
& \qnine{28.48}
\\

proposed: EKF update OFF for $q_{k|k} \geq 0.96$
& \qgtwo{21.21}
& \qgzero{106.06}
& \qgfour{114.85}
& \qhigh{47.97}
& \qhigh{48.20}
& \qhigh{35.05}
\\

\hline

\gls{cfar}
& \qgzero{0.00}
& \qgzero{106.06}
& \qgone{106.06}
& \qgthree{86.62}
& \qsix{78.81}
& \qgfour{85.48}
\\

\gls{cfar}-BC: EKF ON all the time
& \qgzero{0.00}
& \qgzero{106.06}
& \qgthree{84.98}
& \qgzero{106.06}
& \qgzero{106.06}
& \qgtwo{114.13}
\\

\gls{cfar}-BC: EKF update OFF at target arrival
& \qgzero{0.00}
& \qgzero{106.06}
& \qgzero{106.06}
& \qgzero{106.06}
& \qgzero{106.06}
& \qgzero{106.06}
\\

\gls{cfar}-BC: EKF update OFF for $q_{k|k} \geq 0.96$
& \qgzero{0.00}
& \qgzero{106.06}
& \qfive{71.23}
& \qgtwo{85.25}
& \qsix{71.97}
& \qgzero{106.06}
\\

\hline

\end{tabularx}

\vspace{1mm}

\begin{minipage}{0.98\textwidth}
\footnotesize
$^{*}$The target appears at ping index $k=20$.\\
Values with $q_{k|k} \geq 0.96$ are shown in bold.
\end{minipage}

\end{table*}

A similar behavior is observed in the \gls{gospa} values. As the track becomes established, the proposed methods generally achieve lower \gls{gospa} values, reaching approximately $28$--$35$~m at ping $60$ for two of the background-update strategies. The \gls{cfar}-based methods generally retain larger \gls{gospa} values, indicating missed tracks or larger localization errors.

Overall, the proposed likelihood formulation provides more persistent target evidence and generally better localization performance than the \gls{cfar}-based approaches. This improvement results from operating directly on the background-aware raw sensor measurements, allowing weak target evidence to accumulate recursively across pings instead of being discarded by a preliminary detection threshold.
\vspace{-4pt}
\section{CONCLUSION}
To conclude, the results show that explicitly modeling and recursively inferring the background enables effective target tracking in time-varying multipath environments. Incorporating the predicted background statistics into the raw sensor measurement likelihood provides a computationally efficient approach to target tracking without requiring a full joint target--background inference. The evaluation using BELLHOP-generated multipath data further suggests robustness to background-model mismatch while retaining the tracking benefits of the proposed approach. The results also highlight the importance of the background-update strategy in balancing adaptation to temporal background variations against the risk of absorbing target energy into the inferred background statistics. Future work will consider target multipath, stronger model mismatch, multiple-target scenarios, and experimental validation.

\section*{ACKNOWLEDGMENT}
OpenAI ChatGPT was used solely to assist with language polishing and improving the clarity and readability of the manuscript.

\appendices
\section{Wideband linearization of time-varying multipath signal}\label{app:linearization}
Define $F(r,t)\triangleq s(e^r t)$ with $r=\mathrm{ln}\beta$. Then,
\begin{equation*}
    \frac{\partial F(r,t)}{\partial r}= e^r t\dot{s}(e^r t) =t\frac{d F(r,t)}{dt}
\end{equation*}
With $F(0,t)=s(t)$, the solution to the PDE is
\begin{equation}\label{app:eq1}
    F(r,t)=e^{r(t\frac{d}{dt})}s(t), \quad s(\beta t)=e^{\mathrm{ln\beta}(t \frac{d}{dt})}s(t).
\end{equation}
Using the series expansion,
\begin{equation*}
    e^a=1+a +\frac{a^2}{2!}+\frac{a^3}{3!}+\cdots
\end{equation*}
and taking the first order approximation,~\eqref{app:eq1} can be written as
\begin{equation}
    s(\beta t) \approx s(t) +\mathrm{ln}\beta t \dot{s}(t).
\end{equation}

\balance
\bibliographystyle{IEEEtranN}
\bibliography{IEEEabrv, main}

\begin{thebibliography}{28}
\providecommand{\natexlab}[1]{#1}
\providecommand{\url}[1]{#1}
\csname url@samestyle\endcsname
\providecommand{\newblock}{\relax}
\providecommand{\bibinfo}[2]{#2}
\providecommand{\BIBentrySTDinterwordspacing}{\spaceskip=0pt\relax}
\providecommand{\BIBentryALTinterwordstretchfactor}{4}
\providecommand{\BIBentryALTinterwordspacing}{\spaceskip=\fontdimen2\font plus
\BIBentryALTinterwordstretchfactor\fontdimen3\font minus
  \fontdimen4\font\relax}
\providecommand{\BIBforeignlanguage}[2]{{%
\expandafter\ifx\csname l@#1\endcsname\relax
\typeout{** WARNING: IEEEtranN.bst: No hyphenation pattern has been}%
\typeout{** loaded for the language `#1'. Using the pattern for}%
\typeout{** the default language instead.}%
\else
\language=\csname l@#1\endcsname
\fi
#2}}
\providecommand{\BIBdecl}{\relax}
\BIBdecl

\bibitem[Abraham(2019)]{Abraham2019}
D.~A. Abraham, \emph{Underwater {A}coustic {S}ignal {P}rocessing: {M}odeling,
  {D}etection, and {E}stimation}.\hskip 1em plus 0.5em minus 0.4em\relax Cham,
  Switzerland: Springer, Feb. 2019.

\bibitem[Bar-Shalom(2011)]{BarShalom2011}
Y.~Bar-Shalom, \emph{Tracking and data fusion}, P.~K. Willett and X.~Tian,
  Eds.\hskip 1em plus 0.5em minus 0.4em\relax Storrs, CT: YBS Publishing, 2011.

\bibitem[Blackman and Popoli(1999)]{Blackman1999}
S.~S. Blackman and R.~Popoli, \emph{Design and Analysis of Modern Tracking
  Systems}.\hskip 1em plus 0.5em minus 0.4em\relax Boston, US: Artech House,
  1999.

\bibitem[Richards(2022)]{Richards2022}
M.~A. Richards, \emph{Fundamentals of Radar Signal Processing}, 3rd~ed.\hskip
  1em plus 0.5em minus 0.4em\relax New York, USA: McGraw Hill LLC, 2022.

\bibitem[Ristic et~al.(2013)Ristic, Vo, Vo, and Farina]{Ristic2013}
B.~Ristic, B.-T. Vo, B.-N. Vo, and A.~Farina, ``A tutorial on {B}ernoulli
  filters: Theory, implementation and applications,'' \emph{IEEE Trans. Signal
  Process.}, vol.~61, no.~13, pp. 3406--3430, Jul. 2013.

\bibitem[Wang and Jiao(2019)]{Wang2019}
J.~Wang and J.~Jiao, ``Track before detect for low frequency active towed array
  sonar,'' in \emph{Int. Conf. on Signal, Information and Data Processing
  (ICSIDP)}.\hskip 1em plus 0.5em minus 0.4em\relax Chongqing, China: IEEE,
  Dec. 2019.

\bibitem[Buzzi et~al.(2008)Buzzi, Lops, Venturino, and Ferri]{Buzzi2008}
S.~Buzzi, M.~Lops, L.~Venturino, and M.~Ferri, ``Track-before-detect procedures
  in a multi-target environment,'' \emph{IEEE Trans. Aerosp. Electron. Syst.},
  vol.~44, no.~3, pp. 1135--1150, Jul. 2008.

\bibitem[Zhang et~al.(2021)Zhang, Gao, Teng, and Jia]{Zhang2021}
D.~Zhang, L.~Gao, T.~Teng, and Z.~Jia, ``Underwater moving target detection
  using track-before-detect method with low power and high refresh rate
  signal,'' \emph{Applied Acoustics}, vol. 174, p. 107750, Mar. 2021.

\bibitem[Yang(2012)]{Yang2012}
T.~C. Yang, ``Properties of underwater acoustic communication channels in
  shallow water,'' \emph{J. Acoust. Soc. Am}, vol. 131, no.~1, pp. 129--145,
  Jan. 2012.

\bibitem[van Walree(2013)]{Walree2013}
P.~A. van Walree, ``Propagation and scattering effects in underwater acoustic
  communication channels,'' \emph{IEEE J. Oceanic Eng.}, vol.~38, no.~4, pp.
  614--631, Oct. 2013.

\bibitem[Huang et~al.(2013)Huang, Yang, and Huang]{Huang2013a}
S.~H. Huang, T.~C. Yang, and C.-F. Huang, ``Multipath correlations in
  underwater acoustic communication channels,'' \emph{J. Acoust. Soc. Am}, vol.
  133, no.~4, pp. 2180--2190, Apr. 2013.

\bibitem[Li et~al.(2015)Li, Li, Yu, Chen, and Dai]{Li2015}
X.~Li, Y.~Li, J.~Yu, X.~Chen, and M.~Dai, ``{PMHT} approach for multi-target
  multi-sensor sonar tracking in clutter,'' \emph{Sensors}, vol.~15, no.~11,
  pp. 28\,177--28\,192, Nov. 2015.

\bibitem[Li et~al.(2021)Li, Lu, Ali, and Jin]{Li2021}
X.~Li, B.~Lu, W.~Ali, and H.~Jin, ``Passive tracking of multiple underwater
  targets in incomplete detection and clutter environment,'' \emph{Entropy},
  vol.~23, no.~8, p. 1082, Aug. 2021.

\bibitem[Yang et~al.(2024{\natexlab{a}})Yang, Zhang, and Hou]{Yang2024a}
Y.~Yang, B.~Zhang, and X.~Hou, ``Robust cardinalized probability hypothesis
  density filter based underwater multi-target direction-of-arrival tracking
  with uncertain measurement noise,'' \emph{Applied Acoustics}, vol. 216, p.
  109815, Jan. 2024.

\bibitem[Kim(2024)]{Kim2024}
J.~Kim, ``Tracking multiple underwater targets using adaptive {G}aussian
  mixture probability hypothesis density filter with unknown clutter rate,''
  \emph{IEEE Trans. Aerosp. Electron. Syst.}, vol.~60, no.~6, pp. 9154--9162,
  Dec. 2024.

\bibitem[Yang et~al.(2024{\natexlab{b}})Yang, Ling, Sheng, Mu, and
  Jakobsson]{Yang2024}
C.~Yang, Q.~Ling, X.~Sheng, M.~Mu, and A.~Jakobsson, ``Detecting weak
  underwater targets using block updating of sparse and structured channel
  impulse responses,'' \emph{Remote Sensing}, vol.~16, no.~3, p. 476, Jan.
  2024.

\bibitem[Jia and Li(2021)]{Jia2021}
H.~Jia and X.~Li, ``Underwater reverberation suppression based on non-negative
  matrix factorisation,'' \emph{J. Sound Vib.}, vol. 506, p. 116166, Aug. 2021.

\bibitem[Zhu et~al.(2022)Zhu, Duan, and Yang]{Zhu2022}
Y.~Zhu, R.~Duan, and K.~Yang, ``Robust shallow water reverberation reduction
  methods based on low-rank and sparsity decomposition,'' \emph{J. Acoust. Soc.
  Am}, vol. 151, no.~5, pp. 2826--2842, Apr. 2022.

\bibitem[Koul et~al.(2026)Koul, Hendeby, and Skog]{Koul2026}
\BIBentryALTinterwordspacing
A.~Koul, G.~Hendeby, and I.~Skog, ``Tracking time-varying multipath channels
  for active sonar applications,'' in \emph{29th International Conference on
  Information Fusion (FUSION)}, Trondheim, Norway, Jun. 2026, accepted for
  publication. [Online]. Available: \url{https://arxiv.org/abs/2602.15555}
\BIBentrySTDinterwordspacing

\bibitem[Liu et~al.(2012)Liu, Zakharov, and Chen]{Liu2012}
C.~Liu, Y.~V. Zakharov, and T.~Chen, ``Doubly selective underwater acoustic
  channel model for a moving transmitter/receiver,'' \emph{IEEE Trans. Veh.
  Technol.}, vol.~61, no.~3, pp. 938--950, Mar. 2012.

\bibitem[Urick(2013)]{Urick2013}
R.~J. Urick, \emph{Principles of Underwater Sound}, 3rd~ed.\hskip 1em plus
  0.5em minus 0.4em\relax Los Altos, CA: Peninsula Publishing, 2013.

\bibitem[Waite(2002)]{Waite2002}
A.~D. Waite, \emph{Sonar for Practising Engineers}, 3rd~ed.\hskip 1em plus
  0.5em minus 0.4em\relax Chichester, UK: John Wiley \& Sons, Ltd., Mar. 2002.

\bibitem[Stojanovic and Preisig(2009)]{miliucom}
M.~Stojanovic and J.~Preisig, ``Underwater acoustic communication channels:
  Propagation models and statistical characterization,'' \emph{{IEEE} Commun.
  Mag.}, vol.~47, no.~1, pp. 84--89, Feb. 2009.

\bibitem[Bishop(2006)]{Bishop2006}
C.~M. Bishop, \emph{Pattern Recognition and Machine Learning}.\hskip 1em plus
  0.5em minus 0.4em\relax Berlin, Heidelberg: Springer-Verlag, 2006.

\bibitem[Mahler(2007)]{Mahler2007}
R.~P.~S. Mahler, \emph{Statistical multisource-multitarget information
  fusion}.\hskip 1em plus 0.5em minus 0.4em\relax Boston, US: Artech House,
  Feb. 2007.

\bibitem[Porter(2011)]{porter2011bellhop}
M.~B. Porter, ``The {BELLHOP} {M}anual and {U}ser’s {G}uide: {PRELIMINARY
  DRAFT},'' \emph{Heat, Light, and Sound Research, Inc., La Jolla, CA, USA,
  Tech. Rep}, vol. 260, 2011.

\bibitem[Rahmathullah et~al.(2017)Rahmathullah, Garcia-Fernandez, and
  Svensson]{Rahmathullah2017}
A.~S. Rahmathullah, A.~F. Garcia-Fernandez, and L.~Svensson, ``Generalized
  optimal sub-pattern assignment metric,'' in \emph{20th Int. Conf. on
  Information Fusion (Fusion)}.\hskip 1em plus 0.5em minus 0.4em\relax Xi'an,
  China: IEEE, Jul. 2017.

\bibitem[Hasselmann et~al.(1973)Hasselmann, Barnett, Bouws, Carlson,
  Cartwright, Enke, Ewing, Gienapp, Hasselmann, Kruseman,
  et~al.]{hasselmann1973measurements}
K.~Hasselmann, T.~P. Barnett, E.~Bouws, H.~Carlson, D.~E. Cartwright, K.~Enke,
  J.~Ewing, A.~Gienapp, D.~Hasselmann, P.~Kruseman \emph{et~al.},
  ``Measurements of wind-wave growth and swell decay during the joint north sea
  wave project ({JONSWAP}).'' \emph{Ergaenzungsheft zur Deutschen
  Hydrographischen Zeitschrift, Reihe A}, 1973.

\end{thebibliography}

\end{document}